\documentclass[%
 aip,
 jmp,%
 amsmath,amssymb,
preprint,%
a4paper
]{revtex4-2}

\usepackage{graphicx}
\usepackage{dcolumn}
\usepackage{bm}
\usepackage{plimsoll}

\usepackage{xcolor}
\usepackage{mhchem}
\usepackage{colortbl}

\definecolor{white}{rgb}{1,1,1}

\begin{document}

\preprint{AIP/123-QED}

\title[Bromine and Iodine in Atmospheric Mercury Oxidation]{Bromine and Iodine in Atmospheric Mercury Oxidation}

\author{Svend L. Bager}
\author{Luna Zamok}%
 \altaffiliation[Present address: ]{Department of Chemistry, Technical University of Denmark, Kemitorvet Bldg. 207, 2800 Kongens Lyngby, Denmark.}
\affiliation{ 
Department of Chemistry, University of Copenhagen, Universitetsparken 5, 2100 Copenhagen, Denmark
}%

\author{Stephan P. A. Sauer}
 \email{sauer@chem.ku.dk}

\author{Matthew S. Johnson}
 \email{msj@chem.ku.dk}

\date{\today}

\begin{abstract}
We investigate the atmospheric oxidation of mercury Hg(0) by halogens, initiated by Br and I to yield Hg(I), and continued by I, Br, BrO, ClO, IO, \ce{NO2} and \ce{HO2} to yield Hg(II) or Hg(0),
using computational methods with a focus on the creation of data for determining the impact of rising iodine levels. We calculate reaction enthalpies and Gibbs free energies using the Coupled Cluster singlets, doublets, and perturbative triplets method (CCSD(T)) with the ma-def2-TZVP basis set and effective core potential to account for relativistic effects. 
Additionally, we investigate the reaction kinetics using variational transition state theory based on geometric scans of bond dissociations at the CASPT2/ma-def2-TZVP level. 
We compare the results obtained from the CASPT2 and CCSD(T) methods to help define the uncertainty. 
Our results provide insights into the mechanisms of these reactions, and the data produced get us closer to determining iodine's impact on mercury depletion events and on the atmosphere as a whole. The reaction $\cdot$HgBr + Br$\cdot$ → HgBr$_{2}$ was found to be twice as fast as HgI$\cdot$ + I$\cdot$ → HgI$_{2}$, with reaction rate coefficients of 8.8$\times10^{-13}$ and 4.2$\times10^{-13}$ cm$^3$molecule$^{-1}$s$^{-1}$ respectively. The BrHg$\cdot$ + BrO$\cdot$ →  BrHgOBr reaction was about 7.2 times faster than the  $\cdot$HgI + IO$\cdot$ → IHgOI reaction with their rates being 3.3$\times 10 ^{-14}$ and 4.6$\times 10^{-15}$  cm$^3$molecule$^{-1}$s$^{-1}$ respectively. We investigate the Hg$\bullet$XOY (X and Y being halogen) complexes. 
From the reactions investigated including iodine, the reaction with the most plausible chance of impacting the mercury lifetime in the atmosphere is HgI$\cdot$ + I$\cdot$ → HgI$_{2}$. 

%
\end{abstract}

\maketitle

\section{Introduction}

In this paper we use computational methods to investigate the chemical interactions of mercury with bromine and iodine in the atmosphere, focusing in particular on local phenomina and Arctic Mercury Depletion Events (AMDEs). Interest in halogens reacting with mercury is driven by their role in promoting the oxidation of Hg to Hg(II) during AMDEs, leading to mercury depostion to the surface. Research focus has mainly been on the reactions between Br$\cdot$ and Hg. \citep{Dibble2012, jiao2015quality, Horowitz2017} 
It has been suggested that iodine could also play an important role, as an oxidant or by acting to increase the concentration of Br.  \citep{calvert2004potential, Ariya2008} Due to increasing atmospheric iodine concentrations in the Arctic  \citep{saizlopez2012atmospheric}, we have studied the effects of this increase on mercury oxidation and contrast the iodine chemistry with the bromine chemistry. This will be done using computational chemistry to examine the rates of mercury reactions with iodine. 

Mercury persists in the environment long enough for it to be transported all over the globe.  \citep{subir2011review} Over time, mercury pollution from populated areas of the Northern Hemisphere reaches the Arctic. In springtime high levels of halogen radicals can be generated photochemically, and tropospheric ozone in the Arctic is temporarily depleted, and Hg(0) is oxidised.  \citep{steffen2008synthesis} Mercury deposition (and aqueous solubility) increases when it is oxidised, enabling mercury to enter biological ecosystems and rendering them hazardous to humans and animals   \citep{skov2004fate}. 

\bigskip

\subsection{The atmospheric chemistry of mercury}

When entering the atmosphere, the residence time of elemental unoxidized mercury is between 4 and 7 months, which is enough for it to scatter across the Northern Hemisphere.  \citep{saiz2020photochemistry, de2014global, horowitz2017new, holmes2010global} Oxidized mercury, on the other hand, is soluble in water, and its residence time is between hours and days.  \citep{selin2007chemical, Ariya2008} 
If the temperature increases by 10 °C, the amount of evaporated Hg(0) doubles  \citep{kabata2001trace} linking mercury pollution with human activity and climate change.  \citep{kabata2001trace, Xiao1991, Huber2006} 
Mercury can undergo multiple deposition and evaporation cycles furthering it's spread.

Globally the main oxidants for Hg(0) are $\cdot$OH and $\cdot$Br radicals, and the main reactants for the further reaction from Hg(I) to Hg(II) are O$_3$, NO$_2$, and $\cdot$OOH.  \citep{shah2021improved} Shah et al. estimate that $\cdot$Br radicals are responsible for about 75\% of the oxidation of Hg(0) globally and a lifetime of mercury in the atmosphere of 5.5 months. While these are the main reactants, in the Arctic atmosphere during springtime, halogen reactions with unoxidized mercury prove to be even more significant. This is due to emission of halogen compounds and radicals into the atmosphere including from seawater, snow, and ice.  \citep{shah2016origin, schroeder1998arctic, Mason2005, Mason2010} 

In the troposphere, it is estimated that gas phase Hg(II) is the most abundant form, comprising 78\% of Hg(II) mass, while Hg(II) processed by aerosol and cloud droplets comprises 22\%. The total deposition flux of Hg(II) is 5.5 Gg/a, where 71\% of the Hg(II) disposition is situated over the oceans.  \citep{shah2021improved}

A main argument put suggested in previous studies investigating the reaction between Hg and $\cdot$I focus on the  low thermodynamic stability of the Hg(I) complex, leading to rapid decomposition and therefore not contributing to mercury depletion.  \citep{goodsite2004theoretical, cremer2008bonding, auzmendi2014mercury} A similar rationale has been proposed for the reaction between Hg and $\cdot$OH, with similar thermodynamic stabilization energy,  \citep{calvert2005mechanisms} though Dibble et al. find that while the Hg + OH pathway contributes \textless  1\% of the conversion of Hg(0) to Hg(II) globally, it contributes up to 83\% in polluted regions,  \citep{dibble2019modeling} thereby underscoring the complexity of mercury's atmospheric chemistry and its sensitivity to environmental conditions. Though we do not investigate the previous rates calculated for Hg + $\cdot$I $\rightarrow$ HgI$\cdot$, a similar argument as for the reaction between Hg(I) and $\cdot$OH (D$_0$ = 11.02 kcal/mol) could be put forward, as a reaction preceding the reactions of $\cdot$IHg (D$_0$ = 10.03 kcal/mol) with other atmospheric species.

\subsection{Halogen Reactions in the Atmosphere}

Reactive halogen species significantly alter chemical processes in the Arctic atmospheric boundary layer during polar spring.  \citep{simpson2007halogens, steffen2008synthesis} Observations have shown rapid ozone depletion events, with ozone levels dropping from 35$\pm$5 ppb to 10 ppb.  \citep{barrie1988ozone, Oltmans1981} This has been linked to catalytic ozone destruction by halogen atoms.  \citep{barrie1989anthropogenic, iglesias2020natural,villamayor2023very} Notably, studies have found correlations between mercury (Hg) and iodine (I), while bromine (Br) did not show a diurnal cycle.  \citep{spolaor2019diurnal} Although molecular iodine (I$_2$) has not been directly observed in the Arctic, iodine species like IO$\cdot$ have been detected in the Antarctic and sub-Arctic regions.  \citep{Atkinson2012,Rasoa2017}

Approximately 75\% of atmospheric iodine originates from inorganic compounds released from the ocean, with the remainder from organoiodine species.  \citep{carpenter2005abiotic, Prados-Roman2015b} Laboratory studies show that oceanic emissions of hypoiodous acid (HOI) and molecular iodine (I$_2$) are linked to the deposition of tropospheric ozone on the ocean surface, involving reactions with iodide (I$^-$) ions.  \citep{carpenter2013atmospheric, macdonald2014laboratory} This suggests that increased tropospheric ozone and Arctic warming may enhance iodine emissions.  \citep{Prados-Roman2015a, cuevas2018rapid}

\subsection{Ice core measurements indicate rising iodine levels}

Recent studies indicate that iodine chemistry may become more significant as the Arctic warms.  \citep{saizlopez2012atmospheric} Ice core analyses from Greenland show a marked increase in iodine concentrations since 1950, likely due to human influences such as increased tropospheric ozone and Arctic warming.  \citep{cuevas2018rapid} This rise in atmospheric iodine enhances local processing and recycling of iodine in ice and snow.  \citep{saizlopez2015mechanism, kim2016production} Notably, mole fractions of IO up to 1.5 ppt were observed at Alert, Canada.  \citep{Zielcke2015} Particle formation from ppt levels of reactive iodine species like iodic acid (HIO$_3$) has been detected in the Arctic.  \citep{sipila2016molecular} These findings highlight the need to investigate the rates of atmospheric reactions between mercury and iodine to assess the impact of rising iodine levels.

Photochemical reactions in the Arctic saline snowpack involving halogens like Br$_2$, Cl$_2$, and BrCl have been demonstrated both in the field and laboratory.  \citep{Pratt2013, Custard2017} Similar photochemical production of I$_2$ and triiodide (I$_3^-$) from iodide has been shown in Antarctic snow.  \citep{kim2016production} Additionally, iodate (IO$_3^-$) can be photochemically active in frozen solutions.  \citep{Galvez2016} These studies suggest that the photochemical behavior of iodine may be comparable to that of bromine and chlorine in the Arctic. Therefore, investigating the rates of atmospheric reactions between mercury and iodine is crucial to estimate the impact of rising atmospheric iodine concentrations.

In this article, we investigate reactions R1-R3 described below. The reactions are suggested by Goodsite,  \citep{goodsite2004theoretical} Schroeder et al.  \citep{schroeder1998arctic} and Jiao and Dibble.  \citep{jiao2015quality}

\begin{flushleft}
\begin{tabular}{m{1.209cm}m{13.934cm}m{1.257cm}}
~
 &
Hg + M +X$\cdot$ $\rightarrow$ $\cdot$HgX + M &
(R1)\\
~
 &
$\cdot$HgX + Y$\cdot$ + M $\rightarrow$ YHgX +M &
(R2)\\
~
 &
$\cdot$HgX + Y$\cdot$ + M $\rightarrow$ HgXY +M $\rightarrow$ Hg + XY + M &
(R3)\\
\end{tabular}
\end{flushleft}
Here, M is a third body collision partner (N$_2$, O$_2$..), and X$\cdot$ = Br$\cdot$, I$\cdot$, and Y = BrO$\cdot$, ClO$\cdot$, IO$\cdot$, $\cdot$NO$_2$, HO$_2$$\cdot$, I$\cdot$, or Br$\cdot$. The two last reactions were proposed by Jiao and Dibble as likely propagators in the mechanism. In this article, we focus on reaction chains starting from reaction R1 where X = I$\cdot$ and Br$\cdot$.  The results will be compared with results of X = Br$\cdot$ from other calculations and from literature when possible. The goal is to produce thermodynamic and kinetic data so evaluation of the increasing iodine levels local influence on AMDEs can be made possible in the future.


\bigskip

\section{Method}

\subsection{Computational Details}

Reaction energies and rate coefficients were computed using computational quantum chemistry methods. The CCSD(T) method with effective core potentials (ECPs) and the minimally augmented basis set ma-def2-TZVP was employed for calculating reaction energies and properties of mercury at equilibrium geometries, incorporating relativistic effects   \citep{khalizov2003theoretical,balabanov2003mercury,jiao2017first}. These previous theoretical investigations of mercury compounds have shown that this method is good for both geometry optimizations and frequency calculations and has been used for accurate calculations on mercury compounds. This is due to the large contribution from the correlation energy compared to that of the contribution of relativistic effects to the energy, though the latter is still significant. 

Rate coefficients were calculated using variational transition state theory (VTST)   \citep{truhlar1984variational}. Relaxed geometry scans of the dissociation reactions were performed using the CASSCF method with the ma-def2-TZVP basis set to obtain energies and frequencies at intermediate geometries as input for VTST. Due to spin-contamination issues with unrestricted CCSD(T) at intermediate geometries, a multiconfigurational approach was necessary.

Single-point energy calculations were conducted using CASPT2 with the same basis set. Frequency calculations employed either CASPT2 or CASSCF methods with the ZORA relativistic approximation and the ZORA-TZVP basis set, as the ECP approximation resulted in multiple imaginary frequencies. Approximately 47 intermediate single-point energy and frequency calculations were performed. See Figure \ref{fig:discurves}

All electronic structure calculations were carried out using the ORCA program   \citep{ORCA12,ORCA18,ORCA22}. Input files for these calculations can be found in the Supplementary Material Section 3.1.

\subsection{Kinetics}
Variational transition state theory (VTST) was chosen because the mechanisms underlying the reactions between mercury and bromine or iodine radicals have a unique characteristic in that all the investigated reactions only have a reaction barrier when excited rotational and vibrational states are taken into account.
The Multiwell program suite was used to carry out the VTST calculations. \citep{barker2022multiwell} Multiwell uses the subprogram Ktools for unimolecular dissociation reactions with rovibrational reaction barriers taken into account. From Ktools, the high-pressure rate coefficients for the unimolecular dissociation reaction are given by equation (\ref{eqn:2}). The high-pressure rate coefficients the for reverse reaction in association reactions are calculated using the equilibrium constant relation.


\subsubsection{Angular momentum resolved microcanonical VTST in Ktools.}
The formulation used in this study is based on the one used in the Multiwell program package. \citep{barker2022multiwell, Yang2015} All relevant equations are detailed in the manual, primarily in the sections describing the theory behind the Ktools (p. 114) and sctst (p. 209) subprograms. Equation (\ref{eqn:2}) outlines the method for calculating rate coefficients in Ktools 
for uni-molecular dissociation reactions, averaged over the electronic, vibrational and rotational energy at a specific temperature.

\begin{equation}
k\left(T\right)=\frac{\sum _{J=0}^{J_{\mathit{max}}}\int
_{\epsilon=0}^{{\infty}}N^{{\ddag}}\left(\epsilon,J\right)\exp
[-\left(\epsilon+E_{\mathit{VT}}+B_{\mathit{VT}}^{+} {\left(J+1\right)}\right)/k_BT]\mathit{d\epsilon}}{h\sum
_{J=0}^{J_{\mathit{max}}}\int _{\epsilon=0}^{{\infty}}\rho \left(\epsilon,J\right)\exp
[-\left(\epsilon+E_0+B_R^{+} {J\left(J+1\right)}\right)/k_BT]\mathit{d\epsilon}} \label{eqn:2}
\end{equation}

Above, $E$\textsubscript{0} and $E$\textsubscript{VT} denote the potential energies of the reactant and of the variational transition state respectively. The summations are over the angular momentum denoted by quantum number \textit{J} and the integrals are over the energy $\epsilon$. Where $\epsilon$ is used to define a bundle or interval of energy levels of the total rovibrational energy \textit{E\textsuperscript{+}} used in the numeric integration, and it is set as a user-defined parameter in Ktools to increase precision or decrease the size of integrals. \textit{B\textsubscript{VT}} and \textit{B\textsubscript{R}} are the rotational constants for the variational transition state and the reactant, respectively. 

Before using the expression, the reaction path is analysed. Each reaction coordinate on the reaction path (See Figure \ref{fig:discurves}) is treated as a potential transition state. One or more reaction coordinates are identified as local reaction rate minima. If multiple reaction rate minima or bottlenecks are deemed "significant", the transition states are combined to create a "unified rate coefficient" using Miller's unified statistical theory. \citep{Miller1976, Yang2015}

\section{Results}

\subsection{Geometries and Thermodynamics}
We performed geometry optimization and frequency calculations at the CCSD(T) level for the molecules investigated in this study. Reaction geometries and harmonic frequencies can be found in the supplementary material section 1. The Gibbs free energy was determined for each molecule investigated here.

\subsubsection{Reaction R1 with X = Br or I}
 
There is a difference in Gibbs free energy of the reaction between the reaction of Hg+I$\cdot$ (-5.17 kcal/mol) and Hg+Br$\cdot$ (-10.52 kcal/mol). The difference between the two reactions is thus -5.35 kcal/mol. 
This is an important difference since it would favor the formation of $\cdot$HgBr radicals over $\cdot$HgI radicals. 
Geometries and frequencies can be found in the supplementary material section 1.1 Table S1.
 
\subsubsection{Reaction R2 with X = Br or I, and OY= BrO, ClO, IO, or HO$_2$}



For XHgY and HgX$_2$ the largest energy difference between reactant and product is for the HgBr + Br $\rightarrow$ HgBr$_2$ reaction with a reaction Gibbs free energy change of -61.09 kcal/mol.  
The stabilization by the Gibbs free energy change decreases 
from XHgOI over XHgOBr to XHgOCl and XHgOOH. 
Here the reaction Gibbs free energy is always less negative for the IHgOY molecules than the BrHgOY molecules. 
The smallest reaction Gibbs free energy is -25.15 kcal/mol in the case of the IHgOOH reaction. 
There is an energy difference of between 2.65 and 2.70 kcal/mol when interchanging the Br and I at the  $\cdot$ XHg radical in reaction R2. 
There is a general increase in reaction Gibbs free energy with increasing atomic number of the halogen in $\cdot$OY radical of reaction R2. 
The reaction Gibbs energy of the formation of BrHgOOH is 3.0 kcal/mol larger than the  Gibbs free energy for the reaction producing IHgOOH. 
The results for reaction R2 can be seen in Table \ref{tab:2}.

\begin{table*}[h!]
\caption{shows the standard reaction Gibbs free energy,  $\Delta_rG^{\stst}$(298 K), and standard reaction enthalpy, $\Delta_rH^{\stst}$(0 K), for reaction R2 for the reactions involving HgX$_2$ and XHgOY calculated at the CCSD(T)/ECP/ma-def2-TZVP level in this work. X denotes one of the halogens Br or I. All the energies are in kcal/mol.}
\label{tab:2}
\begin{tabular}{lcccc}
\hline
\multicolumn{1}{m{5.15cm}}{\textbf{ Reaction R2}} & \multicolumn{2}{c}{\bfseries X=I} & \multicolumn{2}{c}{\bfseries X=Br}\\
\multicolumn{1}{m{5.15cm}}{\textbf{ }} & 
\multicolumn{1}{m{2.4cm}}{$\Delta_rG^{\stst}$(298 K)} & 
\multicolumn{1}{m{1.99cm}}{$\Delta_rH^{\stst}$ (0 K)} & 
\multicolumn{1}{m{2.46cm}}{$\Delta_rG^{\stst}$(298 K)} & 
\multicolumn{1}{m{2.46cm}}{$\Delta_rH^{\stst}$ (0 K)}\\\hline
X$\cdot$ + XHg$\cdot$ → HgX$_2$ & {}-50.40 & {{}-60.43} & {}-61.09 &  {{}-71.00}\\
XHg$\cdot$ + $\cdot$OI → XHgOI  & {}-50.94 &  {{}-61.63} & {}-53.64 &  {{}-64.24}\\
XHg$\cdot$ + $\cdot$OBr → XHgOBr & -38.76 &  {{}-46.14} & {}-41.41 &  {{}-48.69}\\
XHg$\cdot$ + $\cdot$OCl → XHgOCl & {}-36.06 &  {{}-48.98} & {}-38.69 &  {{}-51.58}\\
XHg$\cdot$ + $\cdot$OOH →XHgOOH & {}-25.15 & {}-36.04 & {}-28.13 & {}-39.00\\
\hline
\end{tabular}
\end{table*}

\subsubsection{Reaction R2 with X = Br or I and Y = NO$_2$}
The X-HgNO$_2$ reaction path exhibits two intermediates and two transition states. 

\begin{center}
    XHg$\cdot$ + NOO $\rightarrow$  XHgNOO  $\rightarrow$  TS1  $\rightarrow$  XHgONO-anti  $\rightarrow$  TS2  $\rightarrow$ XHgONO-syn
\end{center}

\begin{figure}[h!]
\centering
\begin{minipage}{0.18\textwidth}
\includegraphics[width=\linewidth]{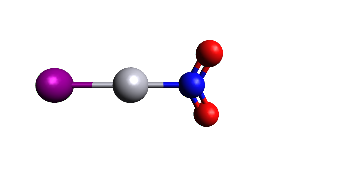}
\end{minipage}
\begin{minipage}{0.18\textwidth}
\includegraphics[width=\linewidth, angle=180, scale={-0.65}]{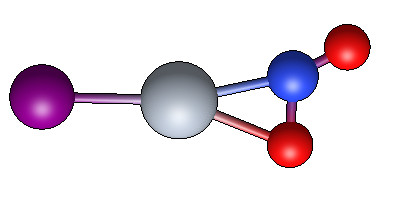}
\hspace{-5mm} 
\end{minipage}
\begin{minipage}{0.18\textwidth}
\includegraphics[width=\linewidth]{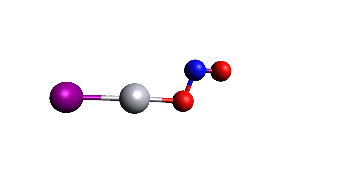}
\end{minipage}
\begin{minipage}{0.18\textwidth}
\includegraphics[width=\linewidth]{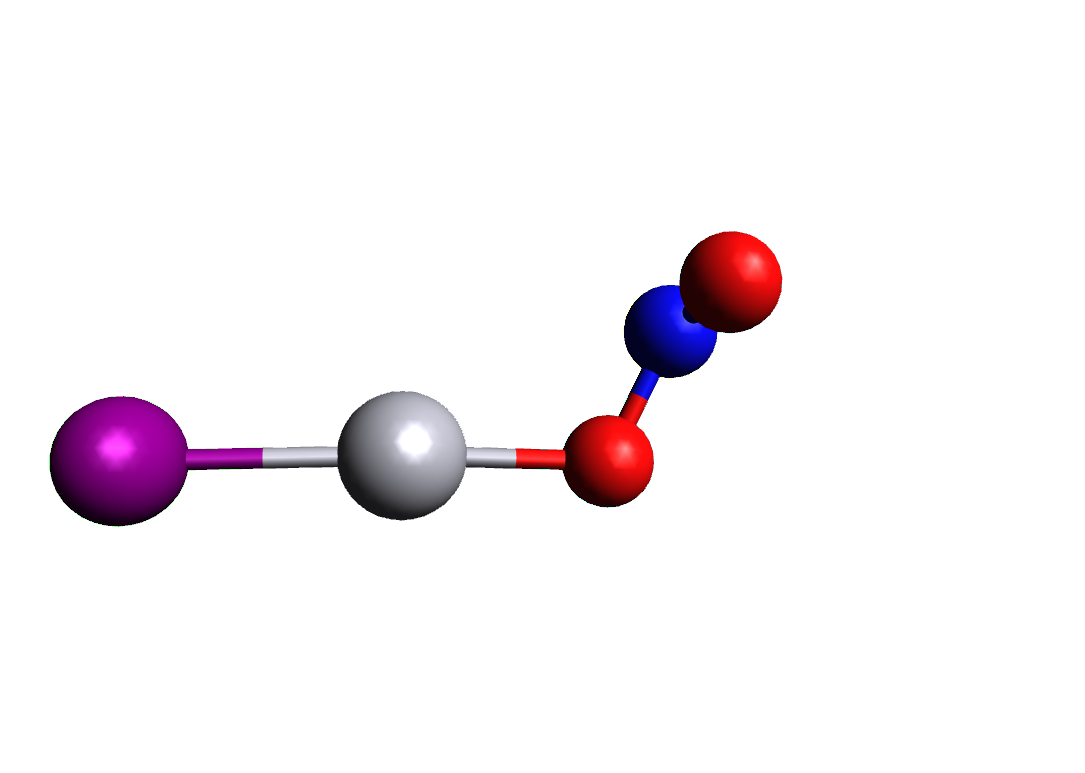}
\hspace{-9mm} \vspace{7mm} 
\end{minipage}
\begin{minipage}{0.18\textwidth}
\includegraphics[width=\linewidth]{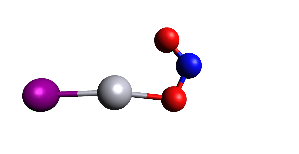}
\end{minipage}
\\
\begin{minipage}{0.18\textwidth} BrHgNO$_2$ \end{minipage} 
\begin{minipage}{0.18\textwidth} TS1 \end{minipage}
\begin{minipage}{0.18\textwidth} BrHgONO-anti \end{minipage}
\begin{minipage}{0.18\textwidth} TS2 \end{minipage}
\begin{minipage}{0.18\textwidth} BrHgONO-syn \end{minipage}
\caption{Show the geometries of the intermediates, transition states and the product for the XHgNOO  $\rightarrow$  TS1  $\rightarrow$  XHgONO-anti  $\rightarrow$ TS2  $\rightarrow$ XHgONO-syn reaction. Starting from the geometry of XHgNOO and ending in XHgONO-syn in the order of the reaction.}
\label{fig:1}
\end{figure}

These can be understood as an example of reaction R2. 
In the first step of this reaction pathway, the lone pair in nitrogen binds to mercury and creates the intermediate XHgNO$_2$. 
Thereafter, a transition state (TS1) creates a barrier for reaching the anti-conformer XHgONO-anti. 
Another transition state (TS2) creates a barrier for reaching the syn-conformer XHgONO-syn, see Figure \ref{fig:1}.

Geometries for products and intermediates have been compared to the result from Jiao and Dibble. \citep{jiao2017first} 
In Jiao and Dibble's article, the geometries of stationary points (reactants, products, and well-defined transition states) and points along the minimum energy path for reactions were optimized using the PBE0 density functional with the aug-cc-pVTZ basis set and relativistic effective core potentials and then verified by harmonic vibrational frequency analyses at the same level of theory. 
The energies of stationary points were refined using the CCSD(T) method in conjunction with the aug-cc-pVTZ basis set. 
In the present study, the geometries were re-optimized, and harmonic frequencies were calculated with the CCSD(T) method, ECPs, and the ma-def2-TZVP basis set.



In this article, the largest difference in bond length of the same configuration is for the halogen mercury bond  X-HgONO with a change in bond distance between 0.16 Å - 0.18 Å and the largest difference in bond angle is 0.9$^\circ$ (XHgO-N-O) for the XHgONO-syn conformers. Further information can be found in the supplementary material section 1.5.

The relative energies for the reaction's IHgNO$_2$ products are similar to those of BrHgNO$_2$, because the geometries along the reaction path are very similar, see Table \ref{tab:3}. 
At TS1 for BrHgONO the energy difference is 12.88 kcal/mol $\Delta_rG^{\stst}$(298 K). 
TS1 for the IHgONO has an energy difference of 12.97 kcal/mol $\Delta_rG^{\stst}$(298 K) when optimized with the CCSD(T) method.
The TS2 energy barrier for the BrHgNO$_2$ and IHgNO$_2$ complex is 5.93 kcal/mol for BrHgNO$_2$ and 7.12 kcal/mol for the iodine complex.

\begin{table*}[h!]
\caption{Shows the standard reaction Gibbs free energy,  $\Delta_rG^{\stst}$(298 K), and standard reaction enthalpy, $\Delta_rH^{\stst}$(0 K), calculated at the CCSD(T)/ECP/ma-def2-TZVP level, of the reconfiguration of IHgNO$_2$ and BrHgNO$_2$ as product of reaction R2.  All the energies are in kcal/mol.}
\begin{tabular}{lcccc}
\hline
\multicolumn{1}{l}{\textbf{Reaction steps}} &
\multicolumn{2}{c}{\textbf{X=I}} &
\multicolumn{2}{c}{\textbf{X=Br}}\\
~
 &
$\Delta_rG^{\stst}$(298 K) &
$\Delta_rH^{\stst}$(0 K) &
$\Delta_rG^{\stst}$(298 K) &
$\Delta_rH^{\stst}$(0 K) \\\hline
 XHg + $\cdot$NOO $\rightarrow$ XHgNOO &
{}-19.42 &
{}-30.87 &
{}-20.58 &
{}-32.80 \\
 TS1 &
{}-6.65 &
{}-18.16 &
{}-7.70 &
{}-19.11 \\
 XHgONO-anti &
{}-20.90 &
{}-32.15 &
{}-22.19 &
{}-34.32 \\
 TS2 &
{}-13.78 &
{}-25.24 &
{}-15.84 &
{}-31.32 \\
 XHgONO-syn &
{}-26.73 &
{}-38.05 &
{}-28.14 &
{}-39.64\\
\hline
\end{tabular}
\label{tab:3}
\end{table*}


\subsubsection{Reaction R3 with X = Br or I and OY = OCl, OBr or OI}
In the article of Jiao and Dibble, \citep{jiao2017first} reaction R3 is suggested as proceeding via BrHgOOH and BrHgONO, occuring as van der Waals complexes. In a similar manner, Hg$\bullet$XOY (X and Y=halogen) complexes are investigated in the present study.



As one would imagine the XHgOY products are more stable than the Hg$\bullet$XOY van der Waals complexes. For the IHgOY and Hg$\bullet$IOY complexes, the difference in Gibbs free energy is around 22 kcal/mol and for the BrHgOY and Hg$\bullet$BrOY complexes the difference is around 31 kcal/mol (Table \ref{tab:5}). 

\begin{table*}[h!]   
\caption{Shows the differences in Gibbs free energy, calculated at the CCSD(T)/ECP/ma-def2-TZVP level, between the XHgOY product and the Hg$\bullet$XOY van der Waals complex in kcal/mol.}
\label{tab:5}
\begin{tabular}{lcc}
\hline
Products – van der Waals complex & \multicolumn{2}{c}{$\Delta $G$^{\stst}$(298 K)} \\
& \multicolumn{1}{c}{\textbf{X=I}} & \multicolumn{1}{c}{\textbf{X=Br}}\\\hline
 XHgOI - Hg$\bullet$XOI   & {}-22.57 & {}-31.76\\
 XHgOBr - Hg$\bullet$XOBr & {}-22.45 & {}-31.52\\
 XHgOCl - Hg$\bullet$XOCl & {}-22.25 & {}-31.24\\
\hline
\end{tabular}
\end{table*}
The area on the potential energy surface surrounding the van der Waals products appears to be a local minimum when looking at the enthalpy.
But when calculating the Gibbs free energy at 298.15 K the local minima changes its characteristics to a saddle point along the minimum energy path. 
This relative shift in the reaction energy of the van der Waals complexes between enthalpy and Gibbs free energy becomes important for the Hg$\bullet$BrOOH and Hg$\bullet$BrNOO complexes that change characteristics from minima to energy barriers, see Table \ref{tab:4}.


The reactions forming the van der Waals complexes (reaction R3), follow the same trend with less negative reaction Gibbs free energies from Hg$\bullet$XOI to Hg$\bullet$XOCl, see Table \ref{tab:4}. However, because of the characteristics of the HgX-OY bond, the Hg$\bullet$BrOY forming reactions have less negative reaction Gibbs free energies than their Hg$\bullet$IOY counterparts.

\begin{table*}[h!]
    \centering
    \caption{Shows the standard reaction Gibbs free energy,  $\Delta_rG^{\stst}$(298 K), and standard reaction enthalpy, $\Delta_rH^{\stst}$(0 K), calculated at the CCSD(T)/ECP/ma-def2-TZVP level, for reaction R3 for the reactions involving Hg$\bullet$XOY.
    All the energies are in kcal/mol.}
\label{tab:4}
%
\begin{tabular}{lccrrr}
\hline\hline
{{\bfseries Reaction R3}} &
\multicolumn{2}{c}{\bfseries X=I } &
\multicolumn{2}{c}{\bfseries X=Br} & \\
~ &
$\Delta_rG^{\stst}$(298 K) &
$\Delta_rH^{\stst}$(0 K) &
$\Delta_rG^{\stst}$(298 K) &
$\Delta_rH^{\stst}$(0 K) \\\hline
{Intermediate - Reactant (van der Waals complex)}\\\hline
{$\cdot$XHg+OI$\cdot$ $\rightarrow$ Hg$\bullet$XOI}   & -28.37 & -38.43 & -21.88 & -31.65\\
{$\cdot$XHg+OBr$\cdot$ $\rightarrow$ Hg$\bullet$XOBr} & -16.30 & -25.97 &  -9.89 & -19.23\\
{$\cdot$XHg+OCl$\cdot$ $\rightarrow$ Hg$\bullet$XOCl} & -13.81 & -23.32 &  -7.45 & -16.65\\
{$\cdot$XHg+$\cdot$OOH $\rightarrow$ Hg$\bullet$XOOH}&  -6.45 & -16.50 &   0.72 & -10.34\\
{$\cdot$XHg+$\cdot$ONO $\rightarrow$ Hg$\bullet$XNO\textsubscript{2}} 
                                             &  -2.06 & -13.18 &   0.66 & -10.94\\\hline
{Product - Intermediate (van der Waals complex)}\\\hline
{Hg$\bullet$XOI $\rightarrow$ Hg+XOI} & -3.93	  & 3.42   & -4.92 & 2.22\\
{Hg$\bullet$XOBr $\rightarrow$ Hg+XOBr} & -3.79	& 3.54	 & -5.79 &	1.11\\
{Hg$\bullet$XOCl $\rightarrow$ Hg+XOCl} & -3.66 &	3.59 & -4.64 & 2.35\\
{Hg$\bullet$XOOH $\rightarrow$ Hg+XOOH}  & -----  & -----  &	-4.54 & 3.38\\
{Hg$\bullet$XNO\textsubscript{2 }$\rightarrow$Hg+XNO\textsubscript{2}}                        
     &  -4.87 & 2.81 & -7.15 & 0.59\\\hline
{Product - Reactant (van der Waals complex)}\\\hline
{$\cdot$XHg+OI$\cdot$ $\rightarrow$ Hg+XOI}    & -32.30 & -35.01 & -26.80 & -29.43\\
{$\cdot$XHg+OBr$\cdot$ $\rightarrow$ Hg+XOBr}  & -20.09 & -22.43 & -15.68 & -18.12\\
{$\cdot$XHg+OCl$\cdot$ $\rightarrow$ Hg+XOCl}  & -17.47 & -19.73 & -12.09 & -14.30\\
{$\cdot$XHg+$\cdot$OOH $\rightarrow$ Hg+XOOH}  & -----   & -----  & -3.82 &  -6.96\\
{$\cdot$XHg+$\cdot$ONO $\rightarrow$ Hg+XNO\textsubscript{2}}
                               &  -6.93 & -10.37 & -6.49 & -10.35\\

\hline\hline
\end{tabular}
%
\end{table*}


\subsubsection{Reaction R3 with X = Br or I and OY = NO$_2$ or OOH}

It was only possible to find the transition states suggested in the article of Jiao and Dibble \citep{jiao2017first}
for the $\cdot$HgBr + $\cdot$OOH $\rightarrow$ Hg$\bullet$BrOOH $\rightarrow$ Hg + BrOOH  reaction (R3) at the B3LYP/ECP/ma-def2-TZVP level of theory, 
but it could not be confirmed at the CCSD(T)/ECP/ma-def2-TZVP level of theory.  
This transition state should create an energy barrier for the enthalpy at 0 K of 1.6 kcal/mol. The reaction enthalpy for the Hg$\bullet$BrOOH "van der Walls" complex was found to be -10.34 kcal/mol compared to Jiao and Dibble’s -11.9 kcal/mol, \citep{jiao2017first} see Table \ref{tab:4}. 
When calculating the Gibbs free energy difference between the intermediate and the reactant it was found to be positive, though only by 0.72 kcal/mol.  
The T1 diagnostics of the Hg$\bullet$BrOOH "van der Walls" complex has a value  of 0.014 (less than 0.02) which suggest that the calculation should be well described by the CCSD(T) single reference method.   

In the case of Hg$\bullet$BrOOH, the van der Waals complex Hg$\bullet$BrNOO has a minimum on the potential energy surface for the enthalpy at 0 K, but when investigating the Gibbs energy at 298 K, the energy of the intermediate this is larger than the energy of the reactant by a small margin of 0.66 kcal/mol, see Table \ref{tab:4}.

Examining the reaction enthalpy at 0 K, it is seen that the van der Waals intermediate  is more stable than the van der Waals product with a reaction barrier to the dissociation product Hg + BrOOH of 3.63 kcal/mol. This changes when compared with the Gibbs free energy in which case the product from the van der Waals complex is lower in energy relative to the van der Waals complex itself.  At 298.15 K the difference for the Gibbs free energy is 4.54 kcal/mol. 

Unlike to the cases of Hg$\bullet$BrOOH and Hg$\bullet$BrNOO, where the dynamics of the reaction are changed from a positive energy difference to a reaction with a negative energy differnence, the Hg$\bullet$INOO and Hg$\bullet$IOOH van der Waals complexes only have negative energy differences, in the case of the reaction enthalpy at 0 K (-16.50 kcal/mol) and the Gibbs free energy at 298.15 K (-6.45 kcal/mol) for Hg$\bullet$IOOH, or in the case of the reaction enthalpy at 0 K (-13.18 kcal/mol) and the Gibbs free energy at 298.15 K (-2.06  kcal/mol) for Hg$\bullet$INOO.   

If reaction R3 for $\cdot$HgX + $\cdot$NO$_2$ was the only reaction considered, then Hg would be more likely to be reformed in the $\cdot$HgBr + $\cdot$NO$_2$ reaction  than in the $\cdot$HgI + $\cdot$NO$_2$ reaction. The same would apply to the $\cdot$HgBr + $\cdot$OOH reaction R3. In the other examples of the reversed reaction R3 investigated, this would also hold true, although the formation of Hg would be significantly more irrelevant and unlikely because of the van der Walls intermediates being significantly more stable.


\subsection{Kinetics}
\begin{figure*}[h!]
    (a)
        \includegraphics[width=0.45\linewidth]{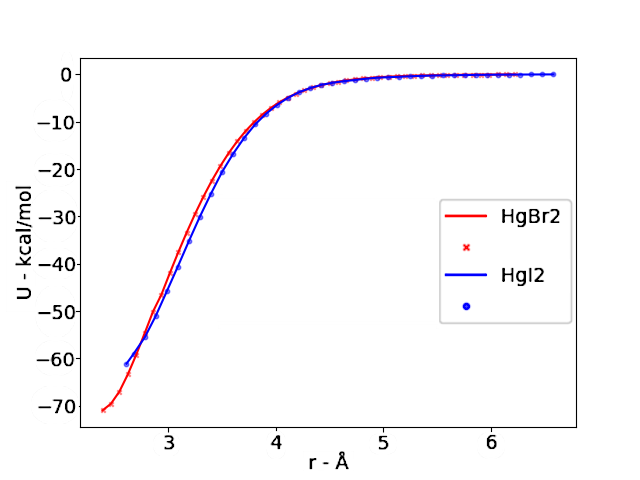}
\\
       (b) \includegraphics[width=0.45\linewidth]{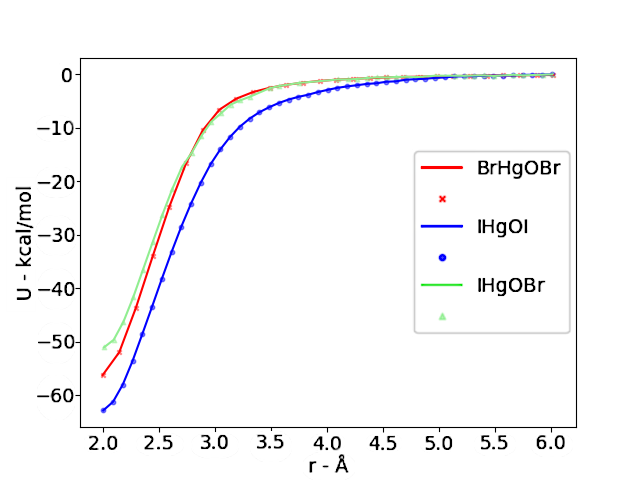}
       (c) \includegraphics[width=0.45\linewidth]{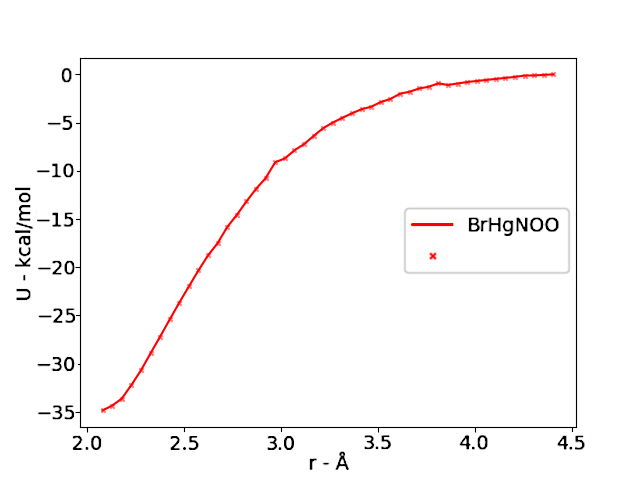}
    \caption{This figure shows the electronic energy in kcal/mol for the dissociation of the bonds of (a) IHg-I and BrHg-Br - (b) BrHgOBr, IHgOI and IHgOBr - and (c) BrHgNO$_2$ calculated with the CASPT2 method. The active spaces used can be found in Table \ref{tab:diffmulten}. The geometry is calculated from CASSCF/ma-def2-TZVP relaxed scans over the dissociating bond.}
    \label{fig:discurves}
\end{figure*}

In the calculation of the reaction rates we have made use of a combination of CASSCF and CASPT2 calculations instead of the CCSD(T) approach employed in the calculation of the reaction energies and Gibbs free energies. Before we will analyse the results of these calculations we will first estimate the accuracy of these calculations by comparing to the CCSD(T) reaction energies. The bond dissociation curves calculated with the CAS methods can be seen in Figure \ref{fig:discurves}.

\subsubsection{Estimation of Precision}
To assess the accuracy of our calculations used to Ktools, we compared the CASPT2 reaction energies obtained from geometric scans with the CCSD(T) reaction energies calculated from reactant and product energies, as shown in Table~\ref{tab:diffmulten}.








\begin{table*}[h!]
\caption{This table shows electronic reaction energies $\Delta E$ for the dissociation reactions calculated with the CCSD(T) and CASPT2 methods, both with ma-def2-TZVP basis sets. Also, the difference in the electronic reaction energies calculated with the CASPT2 and CCSD(T) methods is shown. All energies are in kcal/mol.}
\label{tab:diffmulten}
 
\begin{tabular}{lcccr}
\hline
 {Reaction} & { } & \multicolumn{2}{c}{$\Delta$E} 
 & \multicolumn{1}{c}{Difference $\Delta$E} \\
  {product} &  {active space} & {CASPT2} & {CCSD(T)} & {CASPT2 - CCSD(T)} \\\hline
 {HgBr$_2$} & {12e12o} &   {{}-71.00} &   {{}-71.71} &    {0.71} \\
 {HgI$_2$}  & {10e10o} &   {{}-61.58} &   {{}-60.95} &   {-0.63} \\ 
 {BrHgOBr}  & {10e10o} &   {{}-56.24} &   {{}-54.87} & {{}-1.37} \\ 
 {BrHgOI}   & {12e12o} &   {{}-57.92} &   {{}-66.02} &    {8.10} \\
 {IHgOBr}   & {14e14o} &   {{}-51.03} &   {{}-50.16} & {{}-0.87} \\
 {IHgOI}    & {10e10o} &   {{}-62.81} &   {{}-63.35} &    {0.53} \\
 {BrHgNO$_2$} & {16e13o} & {{}-34.84} &   {{}-34.29} & {{}-0.55} \\
\hline
\end{tabular}
  
\end{table*}

The electronic reaction energies calculated by the CASPT2 method differ only slightly from the CCSD(T) energies. The largest discrepancies are 1.37 kcal/mol for BrHgOBr and 8.10 kcal/mol for BrHgOI. While the former is close to the chemical accuracy of 1~kcal/mol, the latter suggests that the kinetic results for reactions forming BrHgOI may be less reliable. As a rule of thumb, a difference of the order of the chemical accuracy in the activation energy leads to a factor of 5 in the rate coefficient.



Frequency calculations using the ECP relativistic approximation resulted in too many imaginary frequencies, so the ZORA approximation was employed despite increasing computational cost. Direct comparison between ECP and ZORA frequencies is not feasible, in this instance.


To estimate uncertainties due to different relativistic approximations, we compared reaction energies calculated using ECP CASPT2/ma-def2-TZVP and ZORA CASPT2 or CASSCF/ma-def2-TZVP levels of theory. The largest deviation observed was 16\% for IHgOI, indicating potential uncertainties in our calculations (see  Table \ref{tab:difffreq}).


\begin{table*}[h!]
\vspace{-0.5cm}
\caption{shows the difference in electronic reaction energy  $\Delta E$ (in kcal/mol) between the ECP CASPT2 calculations used for the electronic energy in the kinetic calculations and the ZORA CASPT2 or  CASSCF electronic energy used in for the frequency calculation.
}
\label{tab:difffreq}
\begin{tabular}{cccccc}
\hline\hline
{CAS method}&  & ~ & \multicolumn{3}{c}{ $\Delta E$}  \\
 {frequency} & {active space} & {reaction product} & {ZORA} & {ECP} & {\% difference} \\\hline
 {CASPT2} & {12e12o} & {HgBr$_2$} & -70.50 & -71.00 & {1\%}  \\ 
 {CASSCF} &  {10e10o} & {HgI$_2$} & -59.28  & -61.58 & {4\%}  \\
 {CASPT2} &  {10e10o} & {BrHgOBr} & -52.70 & -57.86 & 9\%  \\ 
 {CASSCF} & {14e14o} & {IHgOBr} & -62.15 & -51.03 & {10\%}  \\ 
 {CASPT2} & {10e10o} & {IHgOI} & -52.56 & -62.81 & 16\%  \\
 {CASPT2} & {10e10o} & {BHgOI} & -55.32 & -57.92 & 5\%  \\
 \hline\hline
\end{tabular}
\end{table*}

\subsection{Rate Coefficient Results}

The reaction rates in this study are calculated using the Multiwell program suite and its subprogram Ktools. The energy of the bond-dissociating reactions has been calculated, taking into account the distinct behavior of biradical reactions. The energy for bond-breaking reactions forming biradicals is not described accurately at the CCSD(T) level of theory (reaction R2). 
Therefore, it has been necessary to use CASPT2 to accurately describe the dissociation energy, \citep{Yang2013} although CCSD(T) reaction energies calculated from the equilibrium geometries of the product and reactants have been relied upon for comparison, Table \ref{tab:diffmulten}. More information about the specifics of the CAS calculations can be found in the next section, and the input files for these calculations can be found in the supplementary material section 3.1.

This article focuses on the difference between the reaction rates of iodine and bromine species.
The kinetics of the reactions generally follow the trend as expected from the reaction energies, as can be seen in Table \ref{tab:rate}.
The reactions containing bromine are generally faster.  When comparing the formation of HgBr$_2$ and HgI$_2$ in reaction R2, there is a difference making the reaction forming HgBr$_2$ twice as fast as the reaction forming HgI$_2$. Furthermore, the reaction forming BrHgOBr is about 7.2 times faster than the one forming IHgOI, about 6.5 times faster than that forming IHgOBr and about 10 times faster than that forming BrHgOI.

\begin{table*}[h!]
\centering
\caption{shows the rate coefficients in cm$^3$molecule$^{-1}$s$^{-1}$ for a selection of the dissociation reactions of reaction R2 at infinite pressure. The rates have been calculated using the the program Ktools.  Electronic energies for biradical reaction R2 have been calculated using ECP CASPT2/ma-def2-TZVP, and frequencies have been calculated using ZORA CASPT2 or CASSCF/ZORA-TZVP.}
\label{tab:rate}
\begin{tabular}{lccccccc}
\hline
  Reactions &  & \multicolumn{6}{c}{R2} \\
 {Product} & ~ &{HgBr$_2$}&{HgI$_2$}&{BrHgOBr}&{IHgOI}&{IHgOBr}&{BrHgOI}\\\hline
  {273 K} &~& {7.9$\times$10$^{-13}$}  & {3.9$\times$10$^{-13}$} & {3.4$\times$10$^{-14}$}  & {4.9$\times$10$^{-15}$}  &  {4.6$\times$10$^{-15}$}  & {3.8$\times$10$^{-15}$} \\ 
  {298 K} &~& {8.8$\times$10$^{-13}$}  & {4.2$\times$10$^{-13}$} & {3.3$\times$10$^{-14}$}  & {4.6$\times$10$^{-15}$} &  {5.1$\times$10$^{-15}$}  & {3.3$\times$10$^{-15}$}\\
\hline
\end{tabular}
\end{table*}

\begin{figure}[h!]
\includegraphics[width=0.7\linewidth]{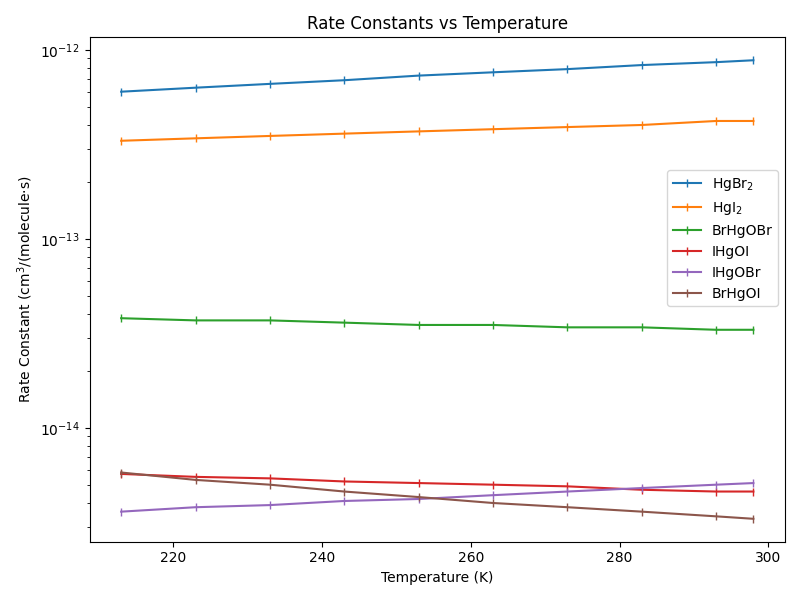}
\caption{show the temperature dependence of the high pressure rate constants described in Table \ref{tab:rate} in cm$^3$molecule$^{-1}$s$^{-1}$. A complete table can be found in the supplementary material section 1.6}
\end{figure}

\begin{table*}[htbp]
\small
\vspace{-1cm}
\renewcommand{\arraystretch}{0.80}
\caption{This table shows rate coefficients from the literature in cm$^{3}$molecule$^{-1}$s$^{-1}$.
}
\label{tab:10}
 
\begin{tabular}{m{4.5cm}m{3.5cm}m{3.552cm}m{3cm}m{1.8599999cm}}
\hline
Reaction &
rate coefficients  &
~ &
Reference &
~ \\\hline 
Hg+I$\cdot$→$\cdot$HgI &
  4.00$\times$10$^{-13}$ &
B3LYP/RRKM &
\cite{goodsite2004theoretical} &
theory\\
Hg + I$_2$ → HgI$_2$  &
  1.27 (±0.58) $\times$10$^{-19}$ &
~ &
\cite{Raofie2008} &
exp. \\
 $\cdot$IHg + I$\cdot$ → HgI$_2$ &{4.2$\times$10$^{-13}$}& CASPT2/VTST & this study & theory\\\hline
Hg+Br$\cdot$→HgBr &
  3.70$\times$10$^{-13}$ &
CCSD(T)/RRKM &
\cite{Goodsite2012} &
theory\\
~  &
  9.8$\times$10$^{-13}$ &
Quas-CT &
\cite{shepler2007hg} &
theory\\
~  &
  6.4(±.2)$\times$10$^{-13}$  &
~
 &
\cite{Sumner2011} &
exp.\\
$\cdot$HgBr+Br$\cdot$→HgBr$_2$ &
  2.50$\times$10$^{-10}$ &
B3LYP/RRKM &
\cite{goodsite2004theoretical}&
theory\\
~  &
  3.0(±0.1)$\times$10$^{-11}$ &
MRCI/Quasi-CT &
\cite{balabanov2005accurate}&
theory\\
~  &
  7.00$\times$10$^{-17}$ &
at 397 K &
\cite{Greig1970}&
exp.\\
 ~ &{8.8$\times$10$^{-13}$}& CASPT2/VTST & this study & theory\\\hline
Hg+Br$_2$→HgBr$_2$ &
  2.8$\times$10$^{-31}$ &
MRCI/VTST &
\cite{balabanov2005accurate}&
theory\\ 
~  &
  9$\times$10$^{-17}$ &
~
 &
  Ariya 2002 \cite{Ariya2002}&
exp.\\
Hg+BrO$\cdot$→HgBrO &
  (3.0–6.4)$\times$10$^{-14}$ &
~
 &
\cite{spicer2002kinetics}&
exp.\\
~  &
  (1–100) $\times$10$^{-15}$ &
~
 &
\cite{Raofie2004}&
exp. \\\hline
$\cdot$HgBr+$\cdot$NO$_2$→BrHgONO &
  8.14$\times$10$^{-11}$ &
PBE0// &
  \cite{jiao2017first}&
theory\\
& & RCCSDT/QuasCT & & \\
Hg+$\cdot$NO$_3$→HgO+$\cdot$NO$_2$ &
  {\textless}4$\times$10$^{-15}$ &
~
 &
\cite{sommar1997gas}&
exp.\\
~  &
  {\textless}7$\times$10$^{-15}$ &
~ &
 \cite{sumner2005environmental}&
exp. \\\hline
Hg+Cl$\cdot$→HgCl$\cdot$ &
  5.40$\times$10$^{-13}$ &
~ &
 \cite{Donohoue2005}&
exp.\\
~  &
  2.80$\times$10$^{-12}$ &
CCSD(T)/CVTST &
\cite{khalizov2003theoretical}&
theory\\
Hg+ClO$\cdot$ → HgClO$\cdot$ &
  1.1$\times$10$^{-11}$ &
~
 &
\cite{byun2010insight}&
exp.\\
~  &
  3.6$\times$10$^{-17}$ &
~
 &
\cite{spicer2002kinetics}&
exp.\\
Hg+Cl$_2$ → HgCl$_2$ &
  2.6(±0.2)$\times$10$^{18}$ &
~
 &
\cite{Ariya2002}&
exp.\\
~  &
  2.5(±0.9)$\times$10$^{-18}$ &
~
 &
\cite{sumner2005environmental}&
exp.\\
Hg + HCl → HgCl$_2$ + H &
  1.50$\times$10$^{-33}$ &
QCISD &
\cite{Wilcox2003}&
theory\\
~ ~ &
  1.00$\times$10$^{-19}$ &
~ &
\cite{Hall1993}&
exp. \\\hline
Hg+F→HgF &
  1.86$\times$10$^{-12}$ &
CCSD(T)/CVTST &
 \cite{khalizov2003theoretical}&
theory \\\hline
Hg+$\cdot$OH→HgOH &
  9.0(±1.3)$\times$10$^{-14}$ &
~ &
  Pal 2004 \cite{pal2004gas}&
exp.\\
~  &
  8.7(±2.8)$\times$10$^{-14}$ &
~
 &
\cite{sommar1997gas}&
exp.\\
HgOH+O$_2$→HgO+OH &
  9.0(±1.3)$\times$10$^{-14}$ &
~ &
\cite{pal2004gas}&
exp. \\\hline
Hg+O$_2$→HgO+O$_2$ &
  7.5(±0.9)$\times$10$^{-19}$  &
~ &
\cite{Pal2004a}&
exp.\\
~  &   3.1$\times$10$^{-40}$ & QCISD(T)//MP2/TST &   Xu 2008 \cite{Xu2008}& theory \\ \hline
Hg + H$_2$O$_2$ → HgO +H$_2$O &
  ${\leq}$4.1$\times$10$^{-16}$  &
~ &
\cite{seigneur1994chemical}&
exp.\\
~
&
  {\textless}8.5$\times$10$^{-19}$ &
293 K &
 \cite{tokos1998homogeneous}&
exp.\\
\multicolumn{2}{l}{$\cdot$HgBr + $\cdot$OOH → BrHgOOH} &
PBE0//&
\cite{jiao2017first}&
theory \\
& 3.51$\times$10$^{-11}$  & RCCSDT/QuasCT & & \\\hline
 BHg$\cdot$ + $\cdot$OBr → BrHgOBr  &{3.3$\times$10$^{-14}$}& CASPT2/VTST & this study & theory\\
  IHg$\cdot$ + $\cdot$OI → IHgOI & {4.6$\times$10$^{-15}$} & CASPT2/VTST & this study & theory\\
\rowcolor{white}   IHg$\cdot$ + $\cdot$OBr → IHgOBr &{5.1$\times$10$^{-15}$}& CASPT2/VTST & this study & theory\\
 \rowcolor{white}   BrHg$\cdot$ + $\cdot$OI → BrHgOI &{3.3$\times$10$^{-15}$}& CASPT2/VTST & this study & theory\\
\hline
\end{tabular}
\end{table*}
\newpage

\section{Discussion} 

\subsection{Comparison with literature rate coefficients}


Of the reactions it has only been possible to compare the rate coefficients calculated by Ktools for reaction R2: Br$\cdot$ + $\cdot$HgBr → HgBr$_2$ with the rate coefficients presented in the literature. The forward rate coefficients given here are calculated by Ktools as the reverse rate coefficient divided by the equilibrium constant. 


When the reverse rate coefficient for reaction R2: HgBr$_2$ → Br$\cdot$ + $\cdot$HgBr  is compared with the one calculated by Balabanov et al. \citep{balabanov2005accurate} (5.5$\times$10$^{-39}$ s$^{-1}$), the reverse rate coefficient calculated in this study (3.5$\times$10$^{-39}$ s$^{-1}$) differs only by a factor of 0.63. 
 The equilibrium constant given in Balabanov's article is 5.407325$\times$10$^{-27}$ molecule$\cdot$cm$^{-3}$, which differs by only a factor of 1.4 from the equilibrium constant calculated in this study 3.91754$\times$10$^{-27}$ molecule$\cdot$cm$^{-3}$. 
 The forward rate coefficient given in Balabanov's article is 3.0$\times$10$^{-11}$ cm$^3$molecule$^{-1}$s$^{-1}$. The difference between the forward rate coefficient calculated in this study (8.8$\times$10$^{-13}$ cm$^3$molecule$^{-1}$s$^{-1}$) and the forward rate coefficient from Balabanov's article would then differ by a factor of 34. However, if a forward rate coefficient is calculated as the reverse rate coefficient divided by the equilibrium constant, both from Balbanov's article \citep{balabanov2005accurate}, the result is 1.017$\times$10$^{-12}$ cm$^3$molecule$^{-1}$s$^{-1}$, which differs from the one calculated by Ktools only by a factor of 1.16. The rate coefficients calculated here are all calculated using the same approach, and are therefore suitable for the comparisons presented here.


The most influential difference of the result in this article is the result of reaction R2: X$\cdot$ + $\cdot$HgX $\rightarrow$ HgX$_2$, both because the rate of the reaction is the fastest calculated in this study and due to the similarities of the rates for the reactions, including bromine and iodine. 
 The rate of reaction R2, including bromine, is close to the rate of the $\cdot$HgBr+$\cdot$NO$_2$→BrHgONO reaction with a rate of 8.1$\times$10$^{-11}$ cm$^3$molecule$^{-1}$s$^{-1}$ and the $\cdot$HgBr+$\cdot$OOH→BrHgOOH reaction with a rate of 3.5$\times$10$^{-11}$ cm$^3$molecule$^{-1}$s$^{-1}$ see Table \ref{tab:diff-rate}.

\begin{table*}[h!]
\caption{ presents the relative difference in reaction rates, calculated by dividing the dissociation rate coefficient of the bromine species on the left-hand side of the table by the dissociation rate coefficient of the iodine species on the right-hand side of the table.
}
\label{tab:diff-rate}
\begin{tabular}{lcllcccr}
 \hline \hline
 {Br}&{Rate} & ~ &{I} & {Rate} & ~ &{Relative} \\
 {species}  &{cm$^3$molecules$^{-1}$s$^{-1}$} & ~ &{species} & {cm$^3$molecules$^{-1}$s$^{-1}$} & ~ &{Rate}  \\
 \hline
 {HgBr$_2$} &
   {8.8$\times$10$^{-13}$} &
 {devided by} &
 {HgI$_2$} &
   {4.2$\times$10$^{-13}$} &
 { =} &
     {2.09}    \\
 {BrHgOBr} &
   {3.3$\times$10$^{-14}$} &
 {devided by} &
 {IHgOI} &
   {4.6$\times$10$^{-15}$} &
 { =} &
     {7.20}    \\
 {BrHgOBr} &
   {3.3$\times$10$^{-14}$} &
 {devided by} &
 {IHgOBr} &
   {5.1$\times$10$^{-15}$} &
 { =} &
     {6.53}    \\
     {BrHgOBr} &
   {3.3$\times$10$^{-14}$} &
 {devided by} &
 {BrHgOI} &
   {3.3$\times$10$^{-15}$} &
 { =} &
     {10.0}    \\
 {IHgOBr} &
   {5.1$\times$10$^{-15}$} &
 {devided by} &
 {IHgOI} &
   {4.6$\times$10$^{-15}$} &
 { =} &
     {1.10}    \\
     {IHgOBr} &
   {5.1$\times$10$^{-15}$} &
 {devided by} &
 {BrHgOI} &
   {3.3$\times$10$^{-15}$} &
 { =} &
     {1.53}    \\
    {IHgOI} &
   {4.6$\times$10$^{-15}$} &
 {devided by} &
 {BrHgOI} &
   {3.3$\times$10$^{-15}$} &
 { =} &
     {1.39}    \\ 

\hline \hline
\end{tabular}
\end{table*}


For reaction R2: (OY$\cdot$ + $\cdot$HgX $\rightarrow$ XHgOY) the rate of the reactions is in the range of 3.3$\times$10$^{-14}$ to 3.3$\times$10$^{-15}$ cm$^3$molecule$^{-1}$s$^{-1}$. 
 The rate coefficient is about 7.2 times higher for reaction R2 including BrHg$\cdot$ and $\cdot$OBr than the one including IHg$\cdot$ and $\cdot$OI. The rate coefficients for reactions forming the products IHgOI, IHgOBr and BrHgOI do not largely differ.

Since the concentrations of OBr$\cdot$ are generally higher than those of I$\cdot$, the BrHgOBr reaction could be significant compared to IHgI formation. For OI$\cdot$ to compete, its concentration must be about 10 times higher than OBr$\cdot$ and up to 18 thousand times higher than those of $\cdot$NO$_2$, $\cdot$OOH, and Br$\cdot$.

Further investigation is needed to understand how iodine competes with bromine in atmospheric reactions, particularly for $\cdot$HgI + $\cdot$NO$_2$ $\rightarrow$ IHgONO and $\cdot$HgI + $\cdot$OOH $\rightarrow$ IHgOOH. While reaction energies favor bromine reactions, kinetic studies are necessary for certainty.

From the literature, Table \ref{tab:10}, there are known reactions, including halogens and mercury that have rate coefficients in a range faster than BrHg$\cdot$ + $\cdot$OBr $\rightarrow$ BrHgOBr (3.3$\times$10$^{-14}$cm$^3$molecule$^{-1}$s$^{-1}$). These are:\\ 
\begin{tabular}{m{1.209cm}m{13.934cm}m{1.257cm}}
~  & Hg + $\cdot$OBr $\rightarrow$ HgOBr & \\
~  & Hg + $\cdot$Cl $\rightarrow$ HgCl & \\
~  & Hg + ClO$\cdot$ → HgClO & \\
~  & Hg + F$\cdot$ → HgF$\cdot$ & \\
\end{tabular}\\
Relevant reactions of the same rate that do not include halogen are the reactions:\\
\begin{tabular}{m{1.209cm}m{13.934cm}m{1.257cm}}
~  & Hg + $\cdot$OH → HgOH & \\
~  & HgOH + $\cdot$O$_2$→HgO(s) + $\cdot$OH & \\
\end{tabular}\\

Since overall reaction rates depend on both rate coefficients and atmospheric concentrations, we provide an estimate of mixing ratios in the local conditions of Arctic and Antarctic in the next section.

\subsection{Atmospheric Mixing Ratios in the Arctic and Antarctic}

High concentrations of bromine, chlorine, and iodine species have been observed during depletion events in both the Arctic and Antarctic. Maximum mixing ratios of BrO$\cdot$ up to 50 ppt in the Antarctic \citep{wagner2007enhanced} and over 20 ppt at Alert Base in the Arctic \citep{Zielcke2015} have been reported. IO$\cdot$ has been measured up to 30 ppt at Mace Head, Ireland \citep{commane2011iodine}, and around 20 ppt at Halley Bay in the Antarctic \citep{saiz2007boundary}. ClO$\cdot$ concentrations reached up to 44 ppt in Barrow, Alaska \citep{custard2016constraints}.

Seasonal variations show that BrO$\cdot$ mixing ratios at Alert Base ranged from 5 ppt in spring to 1 ppt in summer, rising again in fall \citep{Zielcke2015}. At Scott Base in the Antarctic, BrO$\cdot$ increased from below detection limits to about 3 ppt in spring, decreased in summer, and rose slightly in fall. IO$\cdot$ levels at Alert Base varied from 0.1 ppt in spring to 0.3 ppt in summer, while at Scott Base, they ranged from 0.35 ppt in spring and fall to 0.15 ppt in summer.

Other radicals like ClO$\cdot$ and HO$_2\cdot$ were measured during the SOLVE mission over the Norwegian Sea, showing average ClO$\cdot$ concentrations of about 20 ppt in clouds and decreasing from 30 ppt in fall to 10 ppt in winter \citep{simpas2012studies}. HO$_2\cdot$ concentrations dropped from approximately 2 ppt to 0.1 ppt over the same period. Atomic radicals are less frequently reported; Br$\cdot$ concentrations ranged from 4 ppt to 14 ppt during ozone depletion events in Barrow, Alaska \citep{wang2019direct}, and I$\cdot$ was detected up to 22 ppt at Mace Head \citep{bale2008novel}.

Local conditions greatly influence halogen radical concentrations, affecting atmospheric mercury depletion events (AMDEs). While iodine levels may rise due to global warming, current data indicate that bromine radicals, particularly Br$\cdot$ and OBr$\cdot$, are the primary contributors to AMDEs. If iodine species were to become more abundant than bromine, they might compete in atmospheric reactions during mercury depletion events and also play a role in activating bromine \citep{calvert2004potential}. Therefore, although iodine species cannot be dismissed as contributors, the prevalence of bromine radicals suggests they continue to be the dominant factors in AMDEs.

\section{Conclusion and Outlook}
The study conducted provides an analysis and comparison of calculated reaction energies and rate coefficients of mercury's reactions with bromine and iodine, particularly in the context of atmospheric concentrations over the Arctic. 
Utilizing the ORCA quantum chemistry program, the reaction energies were calculated at the CCSD(T) level while the reaction scans were carried out at the CASSCF/CASPT2 level, from which the reaction rates were calculated with the Ktools program. 
CASPT2 calculations of bond dissociation energies were compared to corresponding CCSD(T) energies and the rate coefficients were compared with literature where possible. 

The following observations emerge for the reaction energy. 
The reactions with the largest stabilization of Gibbs free energy creating Hg(II) from reaction R2 ($\cdot$XHg +Y$\cdot$ $\rightarrow$ XHgY or $\cdot$XHg + $\cdot$OY $\rightarrow$ XHgOY) are the reaction forming HgBr$_2$ with a stabilization energy of 61.09 kcal/mol. 

The stabilization by the Gibbs free energy decreases with decreasing atomic number, from XHgOI over XHgOBr to XHgOCl. 
Furthermore, at the other end of the series of molecules, IHgOY molecules are consistently less stable than their BrHgOY counterparts. 
The least stable XHgOY-product is IHgOCl with a stabilization of 36.06 kcal/mol in Gibbs free energy. 

Lower stabilizing reaction Gibbs free energies are observed for the reactions forming XHgOOH and XHgNOO, where the former is more stable than the latter. The reaction forming IHgNO$_2$ has the lowest energy difference among the $\cdot$XHg + $\cdot$OY $\rightarrow$ XHgOY reactions.

The following observations emerge concerning the rate of the reactions.

For reaction R2, the reverse rate and the equilibrium constant were found in agreement with Balabanov et al. \citep{balabanov2005accurate} A dissimilarity was noted in the forward rate coefficient when compared to Balabanov et al. However, the comparison of the rate coefficients in this article should be reliable, as all have been calculated with the same approach.

The most significant difference between iodine and bromine reaction rates was for reaction R2. The bromine-containing forward reaction rate coefficient ($k$= 8.8$\times$10$^{-13}$ cm$^{3}$molecule$^{-1}$s$^{-1}$) was calculated to be twice as fast as its iodine counterpart, with a rate of 4.2$\times$10$^{-13}$ cm$^{3}$molecule$^{-1}$s$^{-1}$.  Calculated reaction rates for reaction R2 ($\cdot$XHg + $\cdot$OY $\rightarrow$ XHgOY ) differed in their forward rate coefficients and were between 3.3$\times$10$^{-14}$ and 3.3$\times$10$^{-15}$ cm$^{3}$molecule$^{-1}$s$^{-1}$. The reaction R2 forming BrHgOBr was about 7.2 times faster than the one forming IHgOI, about 6.5 times faster than the one forming IHgOBr, and about 10 times faster than the one forming BrHgOI.



In accordance of the findings outlined earlier, the reactions between mercury and iodine are not likely to dramatically alter the overall rates of atmospheric reactions involving mercury. This is due to the reactions between mercury and iodine not being faster than those reactions involving bromine, which remains the dominant atmospheric reactant for mercury. Local conditions can occur where iodine concentrations will be dominant, but in the Arctic and Antarctic, the mixing ratio of iodine radical species in the atmosphere is not greater than that of bromine radical species. These observations reinforce the idea that bromine is the primary atmospheric reactant influencing mercury in the Arctic, Antarctic, and in general ADMEs. 
Based on these observations, we can conclude that the AMDEs are dependent on the concentration of these radical species, and the role of iodine species as potential contributors cannot be ruled out. A significant increase in local atmospheric concentrations of iodine would only have a substantial impact on mercury's atmospheric reactions if the atmospheric mixing ratio of iodine markedly surpassed those of atmospheric bromine.
%
%




\end{document}